# A COLLABORATIVE SYSTEM OF SYSTEMS SIMULATION OF URBAN AIR MOBILITY

Nabih Naeem[1] (https://orcid.org/0000-0002-3144-3045),
Patrick Ratei[1] (https://orcid.org/0000-0002-5161-8025),
Prajwal Shiva Prakasha[1] (https://orcid.org/0000-0001-5694-5538),
Lukas Asmer[2] (https://orcid.org/0000-0002-5975-5630),
Roman Jaksche[2] (https://orcid.org/0000-0001-9607-9283),
Henry Pak[2] (https://orcid.org/0000-0002-6259-3441),
Karolin Schweiger[3] (https://orcid.org/0000-0002-8498-9535),
Asija Velieva[3] (https://orcid.org/0009-0001-1130-8399);
Fares Naser[3],
Majed Swaid[4] (https://orcid.org/0000-0002-3017-4680),
Jan Pertz[4] (https://orcid.org/0000-0001-7638-5347),
Malte Niklaß[4] (https://orcid.org/0000-0001-6760-8561)
German Aerospace Center
[1]Institute of System Architectures in Aeronautics, Hamburg
[2]Institute of Air Transport, Köln
[3]Institute of Flight Guidance, Braunschweig
[4]Institute of Air Transport, Hamburg

Contact: Nabih.Naeem@dlr.de

## Abstract

The implementation of Urban Air Mobility represents a complex challenge in aviation due to the high degree of innovation required across various domains to realize it. From the use of advanced aircraft powered by novel technologies, the management of the air space to enable high density operations, to the operation of vertidromes serving as a start and end point of the flights, Urban Air Mobility paradigm necessitates significant innovation in many aspects of civil aviation as we know it today. In order to understand and assess the many facets of this new paradigm, a Collaborative Agent-Based Simulation is developed to holistically evaluate the System of Systems through the modeling of the stakeholders and their interactions as per the envisioned Concept of Operations. To this end, models of vertidrome air-side operations, unmanned/manned air space management, demand estimation and passenger mode choice, vehicle operator cost and revenues, vehicle design, and fleet management are brought together into a System of Systems Simulation of Urban Air Mobility. Through collaboration, higher fidelity models of each domain can be integrated into a single environment achieving fidelity levels not easily achievable otherwise. Furthermore, the integration enables the capture of cross-domain effects and allows domain-specific studies to be evaluated at a holistic level. This work demonstrates the Collaborative Simulation and the process of building it through the integration of several geographically distributed tools into an Agent-Based Simulation without the need for sharing code.

## Keywords

Urban Air Mobility, eVTOL, Agent-Based Simulation, Collaborative Simulation, System of Systems

## NOMENCLATURE

| | |
|---|---|
| UAM | Urban Air Mobility |
| eVTOL | electric Vertical Take-Off and Landing aircraft |
| ABS | Agent-Based Simulation |
| RCE | Remote Component Environment |
| SoS | System of Systems |
| UTM | Unmanned-Air Traffic Manager |
| FATO | Final Approach and Take-Off area |
| MaaS | Mobility as a Service provider |
| GUI | Graphical User Interface |
| CPACS | Common Parametric Aircraft Configuration Schema |

## 1. INTRODUCTION

Urban Air Mobility (UAM) envisions air travel within the urban air space utilizing small aircraft in novel concepts of operation enabled through leveraging advanced technologies and automation. Multiple use cases are classified within the umbrella of Urban Air Mobility such as Intra-City, Mega-City, Airport-Shuttle, Sub-Urban and Inter-City, each with different requirements with respect to technology as well as infrastructure [1]. As UAM presents a significant difference compared to aviation as we know it today, significant research and advancements in fields such as urban air space management, vertiport design and operations, fleet operations, integrated transport, and vehicle design, among the many others (as in FIG 1) are required [2]. In this regard, simulations have been utilized in the literature to carry out investigations with a primary focus on a singular domain. Simulations have been performed in literature to investigate the demand for UAM [3,4], the operating costs [5], concepts for airspace management [6,7], vertiport operations [8–10], as well as vehicle and fleet design [11,12,10,13]. However, a focus on one domain typically comes at the cost of the modelling fidelity of the other domains, which pose a challenge in a UAM network that depends on tight integration between domains. Therefore, not only the advancements in the fields themselves but also the integration between these domains are necessary to enable the envisioned concepts of operations.

As the UAM network is currently still shrouded by significant uncertainty due to its early developmental stage, an integrated modelling of UAM system of systems can serve as a powerful tool in understanding and reducing the uncertainty. From the perspective of a single domain, it may be difficult to evaluate the effect a change in a domain specific parameter may have on the entire UAM network. Consider the number of FATOs at a given vertiport, an analysis solely from the vertiport perspective may be more forgiving to delays due to the limitations of the FATOs during peak hours, but considerations from a holistic perspective may find that the delays at the vertiport in peak hours would reduce the time savings possible for a passenger which may result in passengers rather choosing other modes of transport thereby reducing the revenue realized by the vehicle operator as well as the vertiport operator. Furthermore, delays for landing slots would require the vehicle to loiter in the airspace further congesting the airspace potentially leading to scheduled missions to be delayed if the airspace is saturated, which may in turn impact the revenue of the operators. As demonstrated by this brief example, changes in a single domain may have snowball effects due to the high level of interactions required across these domains, highlighting the need for holistic modelling even when conducting the domain-specific studies.

In order to achieve this holistic modelling, the authors of this work chose to collaborate together to bring together the higher fidelity models of the different domains into a single modelling environment upon which further studies can be performed. A first collaborative approach with various disciplines has been presented by [14], taking a sequential approach to the integration of the disciplines. In this study, a parallel approach is taken to the integration of the disciplines into an Agent-Based Simulation. By employing a collaborative modelling approach, domain specific studies can be carried out by the expert teams in a holistic manner without compromising the fidelity of the other domains.

Agent-Based Modelling was chosen for this integration due to its effectiveness in modelling each stakeholder or entity involved in UAM and the interactions between them as they may occur in a real-life scenario. Through Agent-Based Modelling, the knowledge available to each entity can be modelled through only passing the data that is required/ what would be available to the entity as defined in a concept of operations. This property also makes it suitable to model the On-Demand operations as the foreknowledge of the prospective passenger demand cannot be assumed, which may be more challenging to accomplish through analytical models. Critically, Agent-Based Modelling can be used to model the behavior and decision logic of each stakeholder as independent agents, as they would act in a given scenario. Moreover, Agent-Based Modelling also enables a visual representation of the UAM network which would allow for a better understanding of the operations of the network in addition to the data generated by the simulation. For these reasons, an Agent-Based Simulation was chosen as the backbone upon which all domain tools are integrated. The Hanseatic City of Hamburg was selected as a model city because it is one of the most congested cities in Germany. Furthermore, it is home to a large community of aviation researchers and industry, and it is pushing innovative aviation concepts.

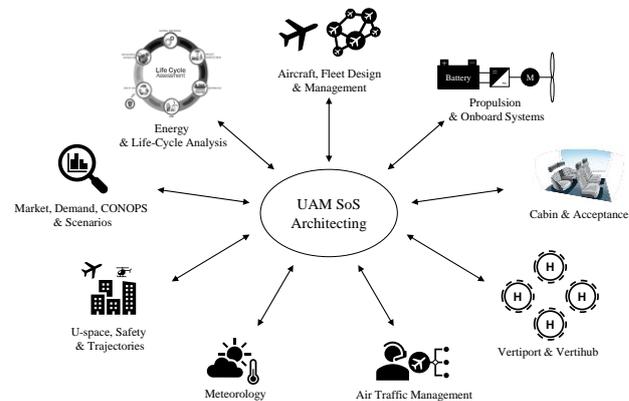

**FIG 1 Urban Air Mobility as a System of Systems, retrieved from [11]**

## 2. METHODOLOGY

Urban Air Mobility can be decomposed into its stakeholders to gain a better understanding of its constituent systems and their interactions. The different stakeholders involved in UAM are considered to be the Customer/Passenger, Mobility as a Service provider, Vehicle Operator (and Manufacturer), Vertidrome Operator, Unmanned-Air Traffic Management Operator (UTM) and the People & Regulators (as in FIG 2). Their roles are considered to be as follows:

1. Vehicle Operator:
   - Operates a fleet of eVTOL vehicles
   - Seeks to maximize profit through transporting the greatest number of passengers at minimal cost
2. UTM & ATM
   - Controls unmanned and manned airspace
   - Seeks to ensure highest level of safety while maximizing operational density of the airspace

3. Vertidrome Operator
   - Operates either one or many Vertidrome(s).
   - Seeks to maximize profit through processing as many passengers as possible in the shortest time
4. Passenger
   - Customer of Urban Air Mobility Service
   - Wants to get from A to B cheaply and quickly
5. Mobility as a Service Provider (MaaS)
   - Operates a platform that connects Passengers with Mobility Services
   - Wants to maximize people using the platform through giving the best mobility services including UAM and connecting ground transport modes
6. People & Regulators
   - Inhabitants, leadership and regulators of the area within which UAM service is provided
   - Wants to ensure safe living environment and minimize disturbances such as noise and visual annoyances

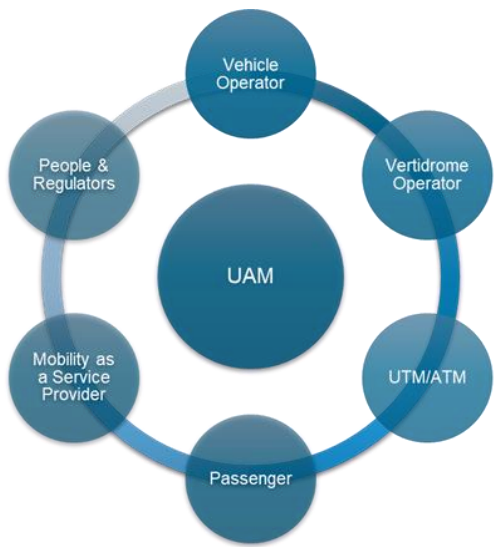

**FIG 2 Urban Air Mobility and its Stakeholders, retrieved from [15]**

## 2.1. Concept of Operations - CONOPS

The Concept of Operations is developed taking inspiration from the user experience of the existing On-Demand ride-sharing services, and the operational requirements from the commercial aviation services and is presented in FIG 3. For UAM to be successful it is envisioned that the Customer interfaces with a singular application/website (MaaS provider), which seamlessly integrates all the aviation service providers and the ground transport options [16], to request travel options and book the desired itinerary from point A to point B with first and last mile options considered. As one of the most significant benefits of UAM is the travel time savings. In order not to lose the time saved it is critical that a seamless integration of UAM with the ground transport modes can be achieved. Once the customer makes the trip request to the MaaS, the MaaS compiles all travel options including ground transport modes, as well as UAM intermodal options together with the price of each option. It is considered that the MaaS communicates with the vehicle operator(s) to find a seat on a suitable vehicle, who in turn communicates with the UTM operator to find a suitable 4D trajectory and departure and arrival slots. The customer, when given the options, can select a binding transport option from those provided. Once the passenger selects the binding offer, the information is passed to the relevant entities and the itinerary is fixed. In this study, it is assumed that there are no deviations from the schedule and that everything happens exactly as planned.

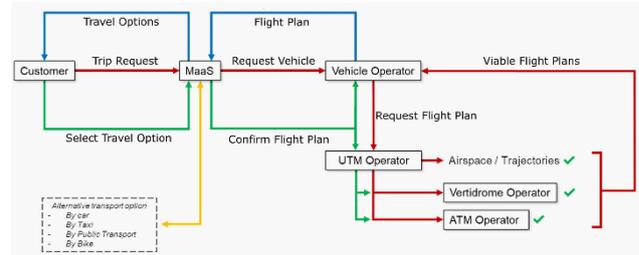

**FIG 3 Booking Process of Trip**

## 2.2. Components of Collaborative Agent-Based Simulation

The stakeholders of UAM are further broken down into domains considering the expertise involved. The passenger is decomposed into the Demand and Mode Choice tools, representing a subset of potential passengers of UAM and a decision model on whether UAM or alternative transport mode is taken respectively. The Vertiport Operator is decomposed into an air and ground side operations, where the airside is represented in this work by the Vertiport Airside Tool. The UTM is represented by the trajectory tool ensuring conflict free routes. The vehicle operator is decomposed into several different disciplines, from the vehicle itself through the Vehicle Design Tool, mission scheduling through the Vehicle Allocation Tool and the Mission Planning Tool, and the economic aspects through the Cost and Revenues Tool. The Agent-Based Simulation is extended to be able to communicate with the individual tools, and this extended version is referred to as the "Agent-Based Simulation Core".

### 2.2.1. Agent-Based Simulation Core

The Agent-Based Simulation Core (ABS-C) is the *agent-based simulation environment* developed in python which serves as the backbone for the integration of the domain-specific tools. The DLR in-house agent-based simulation [17,15,11] is extended in this work to be able to communicate with the other tools, prepare inputs, trigger workflow components, and parse their outputs. The aforementioned communication is achieved through the use of Remote Component Environment (RCE) which allows tool integration and execution as "black boxes", further details can be found in Section 2.3. This extended version of the ABS is referred to as ABS-C in this work. The ABS-C acts as the coordinator and integrator of each tool and triggers each module in the right order as defined by the CONOPS to retrieve the necessary outputs. For each flight request, the ABS-C communicates with the RCE workflow, and after getting its final output, models the customer decision and subsequent flights if any. The ABS-C acts as the active database, as it simulates each stakeholder and their activities throughout the day based on the outputs from the RCE workflow components. In developing the collaborative simulation, it was decided that the ABS-C would act as the single source of truth upon which all domain-specific tools act. This was done to avoid potential accounting errors which could be introduced when dealing with multiple, disconnected databases. At the initialization of the simulation, all air vehicles are distributed across the vertiports as per the user input, are fully charged, and are without any scheduled missions. Throughout the day passengers request and accept flights, flights are created and assigned, passengers are flown, vertiport slots and trajectories are occupied. Consequently, any incoming requests are processed given the exact current state of the UAM network at the time of the request. The GUI representation of the Urban Air Mobility Simulation is depicted in FIG 4, where the considered vertiports in Hamburg, active flights, and passengers at the vertiports can be observed.

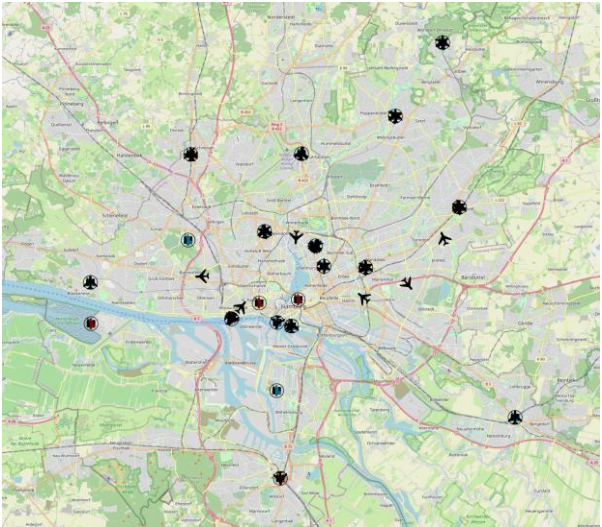

**FIG 4 Visualization of Urban Air Mobility Simulation (base map retrieved from [18,19])**

### 2.2.2. Vehicle Design Tool

The air vehicle which compose the UAM fleet in the Collaborative Simulation are small, vertical take-off and landing capable, fully-electric aircraft. A design tool for vertical take-off and landing aircraft is developed [20] to provide the designs and performance that can be integrated into the Agent-Based Simulation. The design tool is capable of designing vertical take-off and landing aircraft of different architectures including tiltrotor (as in with FIG 5) a broad range of top-level aircraft requirements. Special attention is given here to ensure flexibility of the design tool so that a wide design space can be explored.

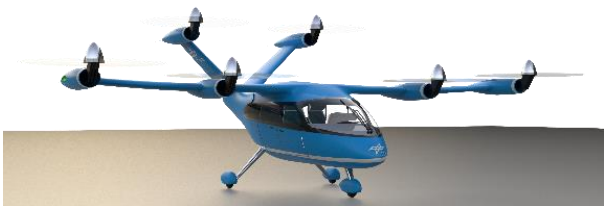

**FIG 5 A 4 Pax Tiltrotor concept. Credits: DLR**

### 2.2.3. Demand Tool

The Demand Tool defines a large dataset of trips in high spatial and temporal resolution (FIG 6 and FIG 7) for which UAM may be a viable option due to the distance between origin and destination.. This dataset contains detailed information about each individual trip including location of origin and destination as well as the time the individual starts their journey. In order to generate travel requests, departure and arrival vertiports that are closest to origin and destination are identified, as well as travel times to them. This allows to determine the earliest time the person would reach the origin vertiport which is needed for mission planning.

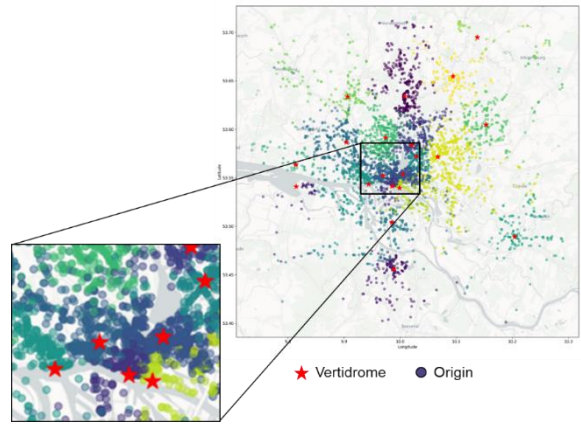

**FIG 6 Spatial distribution of transport demand and vertiports**

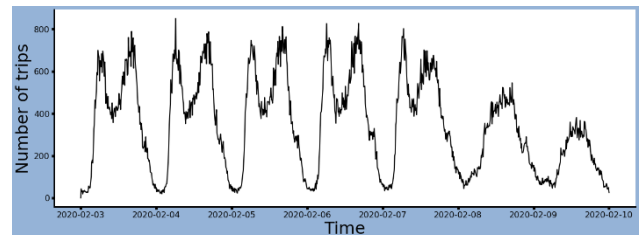

**FIG 7 Temporal distribution of transport demand**

### 2.2.4. Mode Choice Tool

After the actual departure time, flight duration and UAM ticket price are provided by ABS-Core, the Mode Choice Tool first completes the travel chain for the route involving UAM by accounting for first and last mile travel modes. It then constructs a complete route with alternative transport means without the use of UAM. For both complete routes, the time and costs of each mode including the alternative transport modes are considered. Then, the Mode Choice Tool uses a multinomial logit model that uses the total travel times and travel costs of all alternatives as input and determines the probability for each transport mode for each passenger. Based on the calculated probabilities, a random assignment is made to assign the passenger to one of the available means of transport. In case the UAM is selected, the transport offer will be confirmed. In the current version of the simulation, only car is available as an alternative mode, but it is planned to consider other modes at a later stage.

### 2.2.5. Missions Planning Tool

The Missions Planning Tool (component of ABS-C) decomposes a travel request into multi-leg missions if the vehicles in the fleet are unable to perform the direct flight. In the case of a heterogenous fleet, multiple potential mission decompositions are performed and vehicle is searched for each leg. Moreover, for each leg of each mission, a priority is placed on grouping the passenger on existing scheduled missions, only where this is not a possibility, a new mission is requested. Based on the different route options, a final route is chosen for the passenger and each leg in the itinerary is confirmed.

### 2.2.6. Vehicle Allocation Tool

The Vehicle Allocation Tool finds the ideal vehicle from the fleet when a new mission needs to be scheduled considering the existing scheduled missions, the location of the vehicles, available and required energy of the vehicle, and best match between the estimated number of passengers for the mission, and the vehicles passenger capacity. Two separate approaches are available to the Collaborative Simulation in this regard, one through an external tool and the other directly integrated in the ABS-C.

### 2.2.7. Vertidrome Airside Tool

The Vertidrome Airside Tool allocates the next available and conflict-free take-off and landing slot for each UAM request. This requires to keep track of all actual traffic being processed at each vertidrome inside the UAM network. Therefore, a detailed knowledge about the traffic flow and the current airside capacity is necessary. This information is calculated and provided by the V-Lab simulation which describes a discrete-event based simulation covering the airside air and ground operation of a vertidrome and which was specifically adjusted to fit the system-of-systems analysis. The full-size V-Lab simulation module is introduced in [21]. The SoS-tailored V-Lab simulation provides two different vertidrome layouts including two different concepts of operation, on the one hand targeting one-directional independent vehicle traffic flows , and bi-directional interdependent vehicle traffic flows on the other hand [22,8]. Furthermore, it considers designated approach and departure paths for arrival and take-off procedures and performs the overall simulation of the airside traffic flow. A high-level workflow of the Vertiport Airside Tool is displayed in FIG 8.

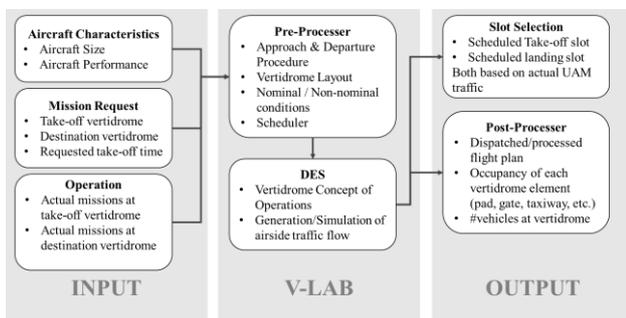

**FIG 8 High-level workflow of the SoS-tailored V-Lab simulation (Vertidrome Airside Tool)**

### 2.2.8. Trajectories Tool

The Trajectories Tool provide a set of trajectories for each mission, based on the origin and destination vertidrome as well as the available time slot at the origin. In order to calculate a conflict-free trajectory, it is necessary to keep track of all the active or planned missions in the airspace. Conflicts are resolved by changing the departure time of the UAM. The calculated trajectories are returned to the Vertidrome Airside Tool with the corresponding arrival time to check, which trajectory matches the available time slot at the destination. The trajectory that satisfies the conditions is then suggested as a possible route. The workflow is summarized in FIG 9. Two route management approaches are defined within the tool, a trajectory-based approach and a slot-based approach. The two approaches are visualized for the city of Hamburg in FIG 10. The slot-based approach considers a fixed route network whereas the trajectory-based approach considers a free route network.

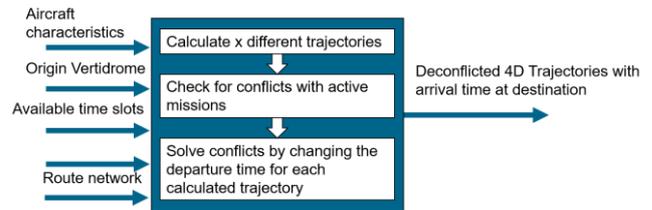

**FIG 9 High-level workflow of the trajectories tool**

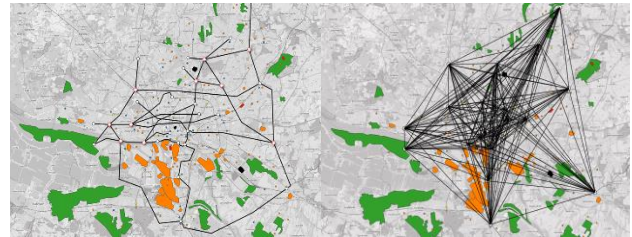

**FIG 10 Airspace management approaches, slot-based approach (left) and trajectory-based approach (right). Retrieved from [23]**

### 2.2.9. Cost and Revenues Tool

The Cost and Revenues Tool sets the base fare and price per kilometer for UAM based on the revenue obtained through ticket sales, and the cost of operation of the fleet. The model considers the fleet size, share of deadhead (empty) flights, energy consumed by the entire network, load factor of the flights, and other parameters in its computations. The ticket price parameters are fine-tuned at the end of each simulation run based on the data from that simulation run. As the ticket price and the mode share achieved by UAM are directly related, the ticket price parameters have to be computed in an iterative loop between the Collaborative Simulation and the Cost and Revenue tool until the ticket price parameters converge.

### 2.3. Integration of Components

A parallel integration of the tools and the Agent-Based Simulation is required to be able to closely model the On-Demand Operations as defined in the CONOPS. Essentially, the tools are to be triggered in a predefined order, using the data from the Agent-Based Simulation Core throughout its runtime i.e. the ABS-C runs in parallel with the intermittently executed tools. This is as opposed to executing each tool and the ABS-C in a sequential manner, one after the other's completion. The developed simulation is governed by the following logic depicted in FIG 11: the demand tool provides a set of trips within Hamburg, and the times at which they depart from their origin and would arrive at the origin Vertiport were they to take a UAM flight. The ABS Core, takes this data as an input, and creates each passenger request to the model a defined time period ahead of their potential arrival at the origin vertiport. By default, this time period is set to 30 minutes, i.e. the passenger requests a UAM flight 30 minutes prior to when they would arrive at the vertiport giving them ample time to make their mode choice selection. This request is processed by the Mission Planning Tool of the ABS-C, which assesses whether the request can be allocated to an already scheduled mission. If this is the case, the Mode Choice Tool is directly triggered and a mode choice decision is obtained. If this is not the case, a new mission must be scheduled, requiring the allocation of the request to a vehicle in the fleet through the vehicle allocation tool which considers the scheduled missions, energy needed and available, and fleet positioning.

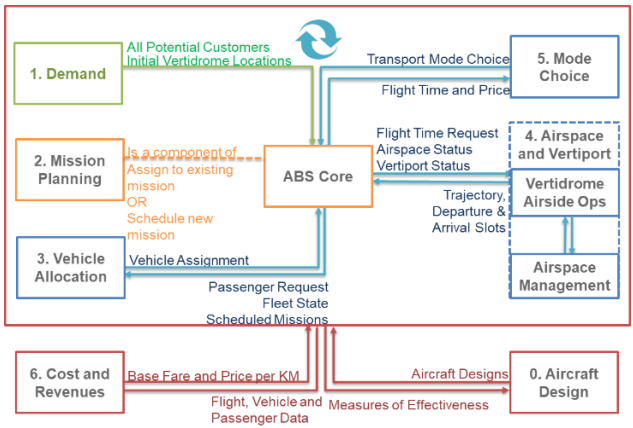

**FIG 11 Flowchart of the collaborative simulation workflow logic**

Once a vehicle is obtained, the vertiport and airspace tools are triggered to find a takeoff FATO slot at the origin vertiport, a trajectory to the destination vertiport, and an arrival FATO slot. Subsequently, the ABS-C compiles the information and based on the ticket pricing structure compiles a total fare and estimated arrival time to the Mode Choice Tool. Based on the chosen mode, the passenger's seat on the flight is fixed. The Vehicle Design Tool provides the aircraft design and performance at the start of the simulation for all vehicle concepts used in the study, which are passed to the tools where needed through the ABS-C. The Cost and Revenues tool is triggered at the end of an entire simulation run, to evaluate and assess the best ticket prices given the history of performed flights.

In order to achieve while still preserving the intellectual property and tool ownership, an integration using Remote Component Environment (RCE) [16] is realized. This approach allows the sharing of tools as "black boxes" through which data is processed, while achieving the desired capabilities without the need for code sharing. Furthermore, using RCE, this tight integration can be achieved in a distributed network where the tools can be located across multiple computers based in multiple locations. The interface between the tools and the ABS-C is set using a version of Common Parametric Aircraft Configuration Schema [24], adapted to describe the Urban Air Mobility network. However, as the tools need to be triggered during the runtime of the Agent-Based Simulation, a typical integration of the Agent-Based Simulation within RCE alongside the other modules would not suffice as a tool cannot stay active within an RCE workflow and communicate with other tools at the same time. The desired behavior is achieved through establishing a link between the ABS and a block within RCE which can be used to transfer the inputs and outputs between RCE and the ABS, this modified version of the ABS is referred to as the ABS-C in this work. Using this approach, the ABS-C can trigger

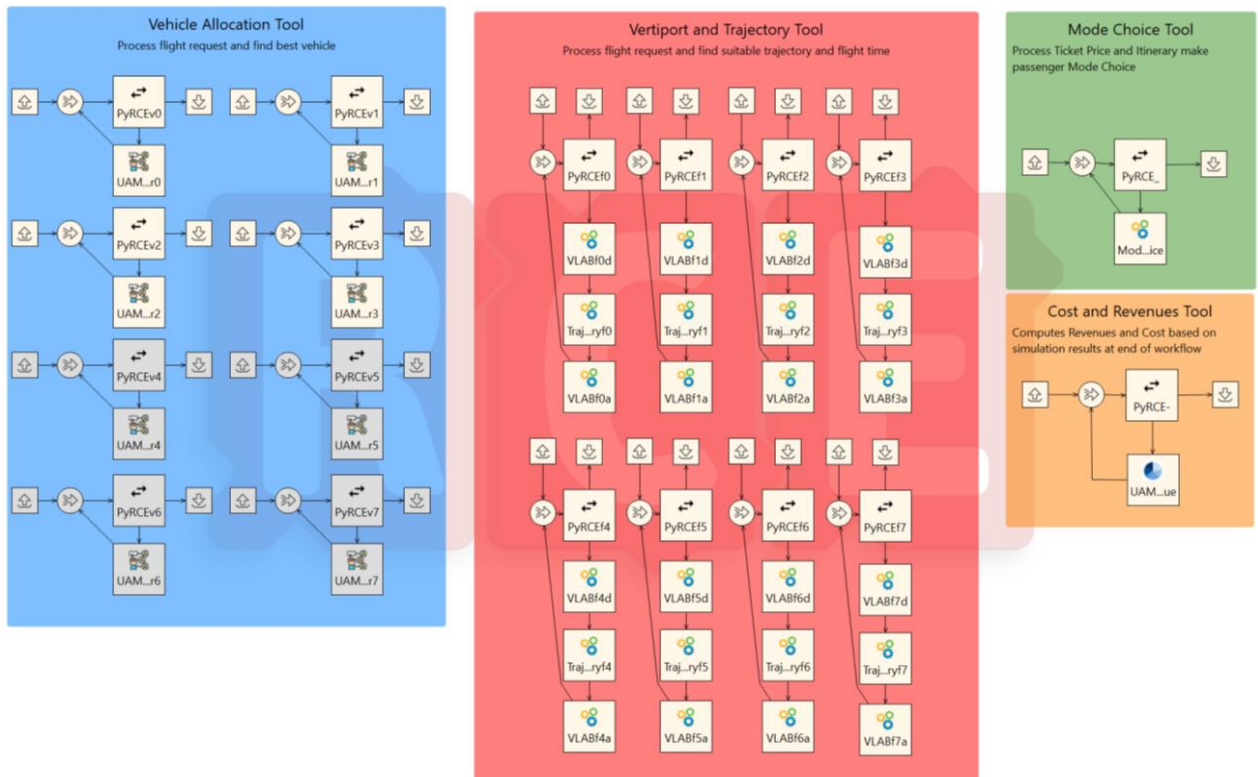

**FIG 12 RCE Workflow for the integration of the tools into the ABS-C, composing the Collaborative Simulation**

```xml
<cpacs>
  <flights>
    <vertiports>
      <vertiport uID="4">
        <vertiportID>4</vertiportID>
        <name>Harburg</name>
        <positionNorth unit="deg">9.7313671</positionNorth>
        <positionEast unit="deg">53.2717517</positionEast>
        <departureTimes>[2006.130101, 2006.130101]</departureTimes>
        <arrivalTimes>[3435.694404, 3435.694404]</arrivalTimes>
      </vertiport>
      <vertiport uID="0">
    </vertiports>
    <requests>
      <request uID="182">
    </requests>
</cpacs>
```

**FIG 13 An excerpt of the interface with ABS and Vertiport and Trajectory Tool**

the workflows in RCE during runtime, passing the relevant inputs and processing the output of each module which in turn are integrated into the ABS-C. The RCE workflow connecting the ABS-C and the domain tools through the PyRCE block is shown in FIG 12. Each domain tool is placed in its own color group with a description, with multiple copies of the tool allowing for parallel processing as detailed in the subsequent section.

The integration of the tools into the ABS-C was achieved through establishing common interfaces. This was done by taking CPACS as the basis, and extending it to be able to describe the UAM system of systems in the fidelity level required. An excerpt of one such interface is depicted on FIG 13. The interfaces were defined with as much commonality as possible, although the flexibility allowed by having the ABS-C as the common interface with all tools meant it was not a requirement.

One major change was made to the overall integration of components to improve the runtime of the simulation: the grouped execution of flight requests. During the development process of the collaborative simulation, it was identified that the computational resources of the tool hosts were not under full utilization during the execution of the Collaborative Simulation with only one tool instance active.

Furthermore, a non-insignificant portion of the overall tool execution time was devoted to the transferring of data between tool hosts and the start-up of tools. Most commonly, RCE workflows are exploited for computationally heavy tools with relatively low iterations needed. In this case however, the computational effort demanded by each tool was minimal however the number of iterations were exceedingly high in the order of thousands, as each passenger request has to be processed through the RCE workflow. As a consequence, this meant that a significant amount of time was dedicated to the transmission of data from one tool to the other over the network. The change implemented by grouping the requests, alleviated both of the aforementioned issues. The logic of the simulation was modified such that the flight requests can be grouped together within a user defined interval and processed at the same time. In order to ensure compatibility with all tools, serial and parallel processing of these grouped requests were implemented. In particular, the vehicle allocation and vertiport and trajectory tools required parallel triggering of the tools for each of the grouped requests due to the specific inputs required. Each flight request in the grouped requests queue is parallelly executed in one of the multiple tool instances in the workflow (see FIG 12). This ensures a higher utilization of the resources available to the tool host, and reduces the time needed for processing the requests by up to the number of tool instances available. For the mode choice tool, serial processing of the grouped requests was implemented, meaning several requests are input to the tool at a time. This reduces the time spent in transferring data and starting up the tools, by reducing the number of times these actions are performed. For the computational resources used in this project, 8 parallel instances of the vertiport and trajectory tool could be realized. However, it is noteworthy to mention that the exact performance improvement achieved is also related to the temporal distribution of flight requests, and the grouping interval used.

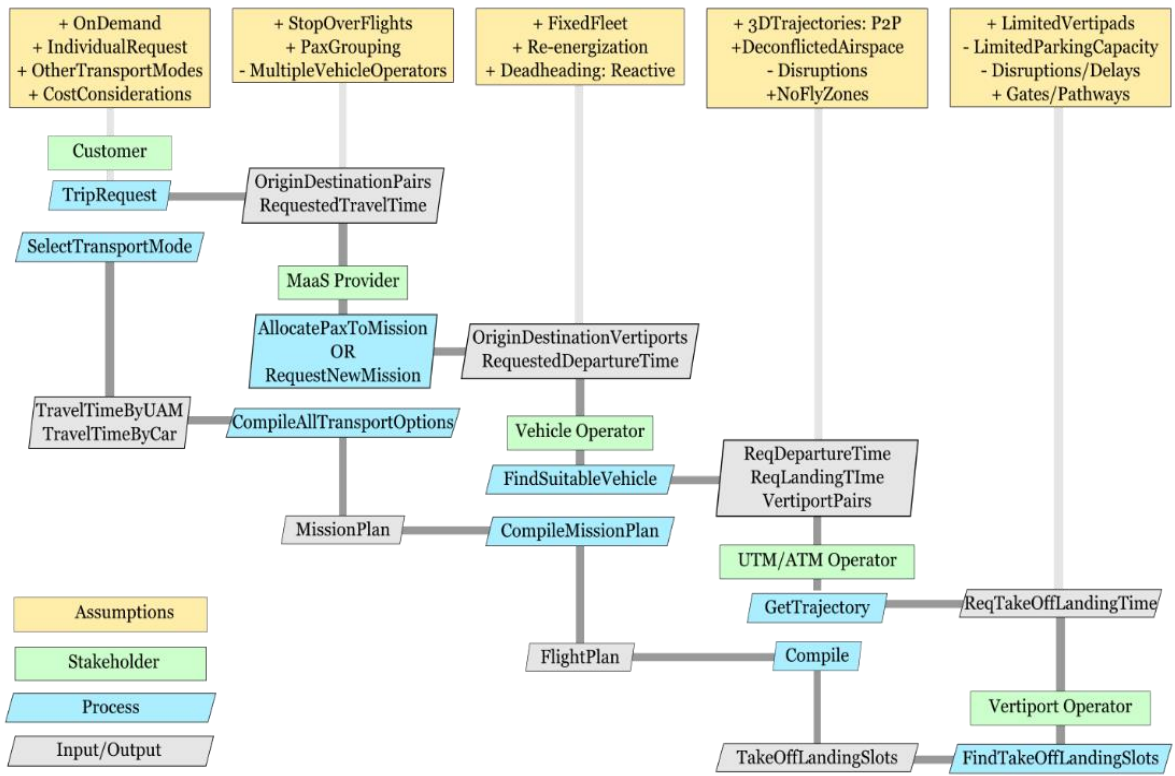

**FIG 14 System of Systems Simulation Design Structure Matrix (S3DSM) for the Collaborative Simulation**

System of systems simulation problems are highly complex, due to the many stakeholders involved and the interactions between these stakeholders. Therefore, a clear dissemination method is needed. Towards this goal: the modified XDSM diagram System of Systems Simulation Design Structure Matrix [15] is presented for the Collaborative Simulation in FIG 14 which clearly states the stakeholders considered, the processes and interactions between them, and the assumptions taken for each stakeholder.

## 3. DISCUSSION AND RESULTS

A study is setup in Hamburg to demonstrate a proof-of-concept of the collaborative simulation. A fleet of 30 vehicles of Tiltrotor Architecture (see FIG 5) with a capacity of 3 passengers and a pilot are distributed across 20 Vertiport locations as defined in FIG 4, and a ticket price of 2 EUR/km is considered. The slot-based approach for airspace management is applied, as depicted in FIG 10. The demand dataset used is as visualized in FIG 6 and FIG 7.

In this work, results are generated for the Hamburg use case over a 4-hour period with several of the modules active within the agent-based simulation, specifically the Vehicle Design Tool, Vertidrome and Trajectory Tools, Mode Choice and the Demand Tools. The Cost and Revenues Tool were not executed in this study, due to the limited window of runtime used, and the Vehicle Allocation Tool was also omitted in this run, in favour of the vehicle allocation algorithm onboard the ABS-C for simplification purposes. In this study, the tools most active within the workflow and without an alternative were included to be able to assess and demonstrate the proof of concept of the collaborative simulation. Future publications will address comparative scenario analyses and sensitivity analyses of the domain-specific parameters.

In order to address the explanatory scope of the collaborative simulation, the results of the brief 4-hour study will be explored. The results are exemplary, and further in-depth investigations are required to make any conclusions with regards to its findings. In the 4-hour window of the study, a total of 1239 flight requests were processed, of which 43 took UAM indicating a 3.4% mode share. The use of the slot-based approach means that longer distances are travelled than the direct point to point distance as with the trajectory-based approach. As the ticket prices are computed based on the actual distance travelled, the slot-based approach indicates higher relative ticket prices according to the assumptions used in this study, which may consequently lead to lower mode share. The mode share of UAM throughout the study period is shown in FIG 15. The distribution of the passengers who chose UAM onto flights and the induced deadhead flights are depicted in FIG 16. It can be observed that the number of flights does not scale proportionally with the passenger, as multiple passengers can be grouped onto the same flight if the conditions are satisfied. The figure also denotes the deadhead flights, which are the non-revenue repositioning flights, required to fulfil the demand when an aircraft is not available at the origin vertiport. These results are of interest to the Vehicle Operator stakeholder. An insight relevant to the UTM stakeholder is given in FIG 17, on the number of aircraft in the airspace at any given time. Such a view can help in understanding whether the airspace is in full utilization, and at which value this is reached.

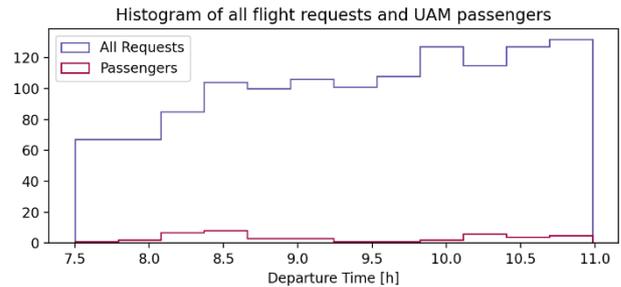

**FIG 15 All flight requests and passengers taking UAM with time**

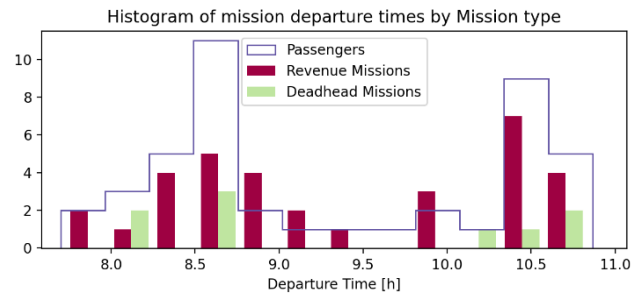

**FIG 16 Passenger and flight times throughout the day**

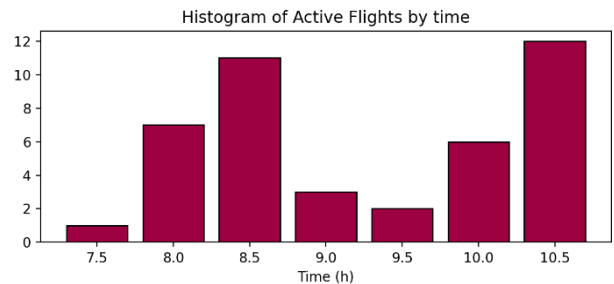

**FIG 17 Number of active in the airspace by time**

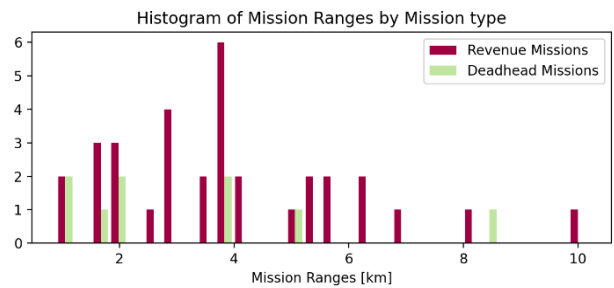

**FIG 18 Distribution of missions by their range**

An insight on ranges of flights for which people are opting for UAM is given in FIG 18. This can be a useful insight for the vehicle operator or designer, and also helps to get a better understanding of the internal functioning of the UAM SoS. Further insights can be derived from the perspective of the passenger, vertiport management, and others. The results presented in this work, are not meant to be conclusive or decisive but rather demonstrating the proof-of-concept of the collaborative simulation. In future work, by evaluating several complete scenarios, deeper insights could be drawn into the UAM network.

## 4. OPPORTUNITIES AND CHALLENGES

The collaborative simulation presents many opportunities for future research. The integration of the domain-specific modules into a SoS simulation enables a broader scope of evaluation. The domain specific studies can be performed and evaluated considering not only its impacts on that specific domain, but also on other domains and the overall SoS. Moreover, as any of the variables within the simulation can be varied, such as vertiport number and placement, separation distances enforced by air traffic management, time taken to clear the FATO area, the uncertainty associated with these domain specific variables can be assessed and better understood with respect to their impact on the overall SoS.

In addition to the evaluation of the impact of domain specific variables on the overall system, complete scenarios can be assessed as well. As an example, low maturity UAM scenario can be compared against a high maturity UAM scenarios in terms of the desired metrics such as mode share achieved, air space occupancy etc.

On the topic of evaluation metrics, in SoS problems it is desirable to evaluate the SoS from the perspectives of each major stakeholder. This is done through definition of key performance indicators (KPIs) at each stakeholder level, which can then be measured based on the specific SoS. The individual stakeholder KPIs could then be merged into a singular KPI by using metrics to evaluate the importance of each stakeholder in the SoS. The Collaborative Simulation can support this as it models the stakeholders and their actions and can collect the relevant data at each stakeholder level to compute their KPIs, and subsequently a singular KPI is so desired.

Another research opportunity is the optimization of the SoS and the constituent systems and domains. The large SoS design space can be explored with the developed collaborative simulation, with the potential for optimization based on a set of criteria and constraints.

The collaborative simulation was developed to simulate nominal conditions, i.e. all constituent systems operate as planned and there are no deviations from the plan. While the nominal conditions are sufficient for addressing many research questions, the off-nominal conditions can open additional avenues of research. As Urban Air Mobility requires operation over inhabited areas at a fast space over densely congested airspace, the investigation of off-nominal conditions is critical to ensure the design of a safe and robust SoS constellation. As the Collaborative Simulation is an Agent-Based Simulation, it is apt for the introduction of stochastic events to model off-nominal conditions and approaches to deal with them. In future work, this field of research can be explored.

When considering the opportunities posed by the collaborative simulation, it is pertinent to keep in mind the challenges associated with it as well. Due to the high degree of collaboration to achieve this, implementing certain changes especially those affecting the interfaces and the logic behind it, can require more effort than a standalone development. In the same way, the execution of studies requires all participating tools to be reliably online for the duration of the run which can exceed half a day. This can often not be the case, and can be a challenge to resolve, more so when the tools are geographically distributed as in the case of the Collaborative Simulation.

Another challenge associated with this distributed approach is the commonly faced issue of debugging. In any software development project, a non-insignificant time contribution goes toward debugging and bug fixing. The time required to debug the collaborative simulation was significantly higher in part due to the nature of collaboration itself and also due to the no code-sharing approach used. In simple terms, bugs can be harder to find when there are more places they could hide in, especially if some of those places are out of sight and reach. At the same time, achieving collaboration without code-sharing is also one of the biggest advantages of this approach, as it ensures the competency is maintained where it was generated. Needless to say, this work is not the only work utilizing RCE workflows to perform studies. However, the unique aspect of this work is due to the integration of the workflow into the Agent-Based Simulation, which provides several benefits but is also the root of many of the challenges faced. As aforementioned, the agent-based simulation acts as the orchestrator, integrator and database for the workflow. This means that a failed workflow execution due to tool unavailability or memory issues, cannot be continued from the failed point and has to be restarted from the same point to ensure consistent and meaningful results. In other words: the simulation is sensitive to its starting point. This coupled with the long runtime and geographical distribution of the tools, can make for a considerable challenge. However, it is of utmost importance to note that many of these issues faced in this work are due to the novel approach taken, and several of the challenges have already been alleviated, and the rest could be entirely solved as the approach is matured.

Overall, the opportunities opened up by this work outweigh the challenges faced, of which many could be resolved. Further efforts can be made to completely resolve the remaining challenges and ensure the ease of exploitation. In this work, the authors were able to demonstrate the proof of concept of the Collaborative Simulation approach taken, and show some of its merits and challenges. The scope for future work is broad, and can take the perspective of any of the individual domains involved in the work, but also most interestingly, the holistic perspective of the UAM network. UAM SoS is shrouded with large degrees of uncertainties and unknowns, such a Collaborative Simulation can aid in the reduction of those uncertainties and unknowns, through integration and assessment of the large body of research in each of the constituting domains of the UAM SoS.

## COMPETING INTERESTS

Henry Pak is also a guest editor for the special issue on the HorizonUAM project but has not been involved in the review of this manuscript.

# References

[1] Asmer, L., Pak, H., Shiva Prakasha, P., Schuchardt, B. I., Weiand, P., et al., "Urban Air Mobility Use Cases, Missions and Technology Scenarios for the HorizonUAM Project," *AIAA Aviation 2021 Forum,* Virtual Event, 2021. doi: 10.2514/6.2021-3198

[2] Schuchardt, B. I., Becker, D., Becker, R.-G., End, A., Gerz, T., et al., "Urban Air Mobility Research at the DLR German Aerospace Center – Getting the HorizonUAM Project Started," *AIAA Aviation 2021 Forum,* Virtual Event, 2021.

doi: 10.2514/6.2021-3197

[3] Fu, M., Straubinger, A., and Schaumeier, J., "Scenario-Based Demand Assessment of Urban Air Mobility in the Greater Munich Area," *Journal of Air Transportation*, Vol. 30, No. 4, 2022, pp. 125–136.
doi: 10.2514/1.D0275

[4] Ploetner, K. O., Al Haddad, C., Antoniou, C., Frank, F., Fu, M., et al., "Long-term application potential of urban air mobility complementing public transport: an upper Bavaria example," *CEAS Aeronautical Journal*, Vol. 11, No. 4, 2020, pp. 991–1007.
doi: 10.1007/s13272-020-00468-5

[5] Pertz, J., Niklaß, M., Swaid, M., Gollnick, V., Kopera, S., et al., "Estimating the Economic Viability of Advanced Air Mobility Use Cases: Towards the Slope of Enlightenment," *Drones*, Vol. 7, No. 2, 2023, p. 75.
doi: 10.3390/drones7020075

[6] Bosson, C. and Lauderdale, T. A., "Simulation Evaluations of an Autonomous Urban Air Mobility Network Management and Separation Service," *2018 Aviation Technology, Integration, and Operations Conference*, Atlanta, GA, USA, 2018.
doi: 10.2514/6.2018-3365

[7] Pinto Neto, E. C., Baum, D. M., Almeida, J. R. de, Camargo, J. B., and Cugnasca, P. S., "A Trajectory Evaluation Platform for Urban Air Mobility (UAM)," *IEEE Transactions on Intelligent Transportation Systems*, Vol. 23, No. 7, 2022, pp. 9136–9145.
doi: 10.1109/TITS.2021.3091411

[8] Schweiger, K., Knabe, F., and Korn, B., "An exemplary definition of a vertidrome's airside concept of operations," *Aerospace Science and Technology*, 2021.
doi: 10.1016/j.ast.2021.107144

[9] Schweiger, K., Knabe, F., and Korn, B., "Urban Air Mobility: Vertidrome Airside Level of Service Concept," *AIAA Aviation 2021 Forum*, Virtual Event, 2021.
doi: 10.2514/6.2021-3201

[10] Swaid, M., Pertz, J., Niklaß, M., and Linke, F., "Optimized capacity allocation in a UAM vertiport network utilizing efficient ride matching," *AIAA AVIATION 2023 Forum*, AIAA AVIATION 2023 Forum, San Diego, CA and Online, American Institute of Aeronautics and Astronautics, Reston, Virginia, 2023.
doi: 10.2514/6.2023-3577

[11] Shiva Prakasha, P., Naeem, N., Ratei, P., and Nagel, B., "Aircraft architecture and fleet assessment framework for urban air mobility using a system of systems approach," *Aerospace Science and Technology*, Vol. 125, 2022.
doi: 10.1016/j.ast.2021.107072

[12] Shiva Prakasha, P., Ratei, P., Naeem, N., Nagel, B., and Bertram, O., "System of Systems Simulation driven Urban Air Mobility Vehicle Design," *AIAA Aviation 2021 Forum*, Virtual Event, 2021.
doi: 10.2514/6.2021-3200

[13] Ratei, P., Naeem, N., and Shiva Prakasha, P., "Development of a UAM Vehicle Family Concept by System of Systems Aircraft Design and Assessment," 2022.

[14] Niklaß, M., Dzikus, N., Swaid, M., Berling, J., Lührs, B., et al., "A Collaborative Approach for an Integrated Modeling of Urban Air Transportation Systems," *Aerospace*, Vol. 7, No. 5, 2020, p. 50.
doi: 10.3390/aerospace7050050

[15] Naeem, N., Ratei, P., and Shiva Prakasha, P., "Modelling and Simulation of Urban Air Mobility: An Extendable Approach," *12th EASN International Conference*, Barcelona, 2022.

[16] Menichino, A., Di Vito, V., Dziugiel, B., Liberacki, A., Hesselink, H., et al., "Urban Air Mobility Perspectives Over Mid-Term Time Horizon: Main Enabling Technologies Readiness Review," *2022 Integrated Communication, Navigation and Surveillance Conference (ICNS)*, 2022 Integrated Communication, Navigation and Surveillance Conference (ICNS), Dulles, VA, USA, 05-Apr-22 - 07-Apr-22, IEEE, 2022, pp. 1–13.
doi: 10.1109/ICNS54818.2022.9771511

[17] Kilkis, S., Shiva Prakasha, P., Naeem, N., and Nagel, B., "A Python Modelling and Simulation Toolkit for Rapid Development of System of Systems Inverse Design (SoSID) Case Studies," *AIAA Aviation 2021 Forum*, Virtual Event, 2021.
doi: 10.2514/6.2021-3000

[18] OpenStreetMaps, "OpenStreetMaps," URL: https://www.openstreetmap.org/.

[19] Bennett, J., *OpenStreetMap. Be your own cartographer / Jonathan Bennett*, Packt Pub, Birmingham, U.K., 2010.

[20] Ratei, P., "Development of a Vertical Take-Off and Landing Aircraft Design Tool for the Application in a System of Systems Simulation Framework," Master Thesis, 6 Jun 2022.

[21] Schweiger, K. and Knabe, F., *Vertidrome Airside Level of Service: Performance-Based Evaluation of Vertiport Airside Operations*, 2023.

[22] Leib, S., "Ausarbeitung und Implementierung eines UAM Vertiport Layout Designs mit zugehörigem Betriebskonzept in eine Vertiportsimulation," Bachelor, 2023.

[23] Schweiger, K., Naser, F., Rabeb, A., and Stasicka, I., "Ground and Air Considerations for Urban Air Mobility Passenger Transport Operations," *Ground and Air Considerations for Urban Air Mobility Passenger Transport Operations*.

[24] Alder, M., Moerland, E., Jepsen, J., and Nagel, B. (eds.), *Recent Advances in Establishing a Common Language for Aircraft Design with CPACS*, 2020.